\documentclass[final,1p,times,twocolumn]{elsarticle}

\usepackage{amsmath}
\usepackage{amssymb}
\usepackage{hyperref}
\usepackage{tensor}

\journal{Physics Letters B}

\begin{document}

\begin{frontmatter}

\title{Negative radiation pressure in the abelian Higgs model}

\author[first]{Dominik Ciurla}
\ead{dominik.ciurla@doctoral.uj.edu.pl}
\affiliation[first]{organization={Institute of Theoretical Physics, Jagiellonian University},
            addressline={Lojasiewicza 11}, 
            city={Krakow},
            postcode={30-348}, 
            country={Poland}}

\begin{abstract}
Interactions of small-amplitude monochromatic plane waves with domain walls in (1+1) dimensional abelian Higgs model with a sextic potential were studied. The effective force exerted on a domain wall was derived from a linearized equation and compared with numerical simulations of the original model. It was shown that the domain walls always accelerate in one direction, regardless of the direction of the incoming wave. This implies that in some cases an effect called negative radiation pressure is observed, i.e. instead of pushing, the wave pulls the soliton.
\end{abstract}


\end{frontmatter}

\section{Introduction}
Solitons are a subject of extensive experimental and theoretical studies in many areas of physics, such as condensed matter physics \citep{Bishop_1979}, nuclear physics \citep{manton2022skyrmions}, Bose--Einstein condensates \citep{Frantzeskakis_2010}, nonlinear optics \citep{malomed2002variational}, cosmology \citep{Kibble_1976}, particle physics \citep{PhysRevD.95.035003} and more. Kinks \citep{vachaspati_2006}, also called domain walls \footnote{Strictly speaking, a kink is a soliton in (1+1) dimensional spacetime, while a domain wall is a higher-dimensional object \citep{vachaspati_2006}. However, since we are interested in interactions of a domain wall with waves perpendicular to it, and neglect its shape across other directions, we can treat the domain wall as a kink embedded in more dimensions and use the names interchangeably.}, are one of the simplest (1+1) dimensional topological solitons \citep{manton_sutcliffe_2004}, existing in scalar field theories such as sine-Gordon,  $\phi^4$, $\phi^6$, double sine-Gordon \citep{CAMPBELL1986165}, Christ--Lee \citep{PhysRevD.12.1606} and many other models. Due to their simplicity, they are often studied in various scenarios, e.g. kink-antikink collisions \citep{CAMPBELL1986165, 10.1143/PTP.61.1550, CAMPBELL19831, doi:10.1137/050632981, PhysRevD.103.045003, PhysRevLett.127.071601, PhysRevD.105.065012, PhysRevE.108.024221}. Domain wall solutions are also present in some gauge theories, especially in (1+1) dimensional version of the Abelian Higgs model \citep{KATSIMIGA2015117, PhysRevD.18.1154, FORGACS1984397, PhysRevD.51.5889}. In this paper, we use this model, albeit with a sextic potential instead of a usually used quartic potential, in order to research the interaction of domain walls with a radiation consisting of small-amplitude plane waves.

Usually, when a plane wave is scattered on a soliton, it pushes the soliton in the direction of propagation: we refer to this as the \textit{positive radiation pressure} (PRP). However, it has been shown that, in some systems, the incoming plane wave can exert a pulling force on the soliton. This effect is called a \textit{negative radiation pressure} (NRP) and was found in many models, including kinks in $\phi^4$ model \citep{Romanczukiewicz:2003tn, Forgacs:2008az}, domain walls in $\phi^6$ model \citep{ROMANCZUKIEWICZ2017295, PhysRevLett.107.091602}, vortices in Gross--Pitaevskii model \citep{PhysRevD.88.125007}, dark-bright solitons in coupled nonlinear Schr\"odinger equations \citep{PhysRevE.109.014228} and other \citep{Romanczukiewicz:2008hi, YAMALETDINOV2019253}. Different mechanisms are responsible for this effect. For example, in the $\phi^4$ model the nonlinear coupling creates a double frequency wave, which, combined with the reflectionlessness in the linear order, exerts NRP. However, a more common mechanism involves transfer of energy from slower to faster modes mainly in the linear order, due to two different dispersion relations of these modes. This can be realized by using two coupled fields (see \citet{PhysRevE.109.014228, Romanczukiewicz:2008hi}), non-topological solitons rotating in phase (such as Q-balls), or by asymmetry of the solitons (see \citet{ROMANCZUKIEWICZ2017295}). The latter mechanism is responsible for NRP in the model studied in this paper: in such a case, waves exert PRP or NRP, depending on the side from which they come.

In this letter, we study NRP exerted on a domain wall. We consider monochromatic plane waves in the gauge field, oscillating only in one direction orthogonal to the direction of propagation, which allows us to reduce the problem to (1+1) dimensions. The incoming waves are treated as linear perturbations of the domain wall, which gives a good description, provided their amplitude is sufficiently small. The slowly moving domain wall, on the other hand, is approximated as a Newtonian particle. Effective force derived from these assumptions can then be compared with numerical solutions of a partial differential equation (PDE) describing the full theory.

The paper is organized as follows. In Sec. \ref{sec:model} the considered model, its domain wall solution, and the conserved charges are presented. Then, in Sec. \ref{sec:linearized} small perturbations of the domain wall are considered, and the $S$-matrix for scattering on the soliton is computed. These results are applied in Sec \ref{sec:dynamics} to compute the effective force exerted by a radiation on a domain wall. Finally, the effective force is compared with the numerical simulations in Sec. \ref{sec:results}.

\section{The model}
\label{sec:model}
Let us consider the Higgs model, i.e. a complex scalar field $\phi$ in (3+1) dimensions coupled with an abelian gauge field $A_\mu$:
\begin{equation}
\label{eq:lagrangian}
\mathcal{L} = - \frac14 F_{\mu \nu} F^{\mu \nu} + (D_\mu \phi)^* D^\mu \phi - U(|\phi|^2) \,,
\end{equation}
where $D_\mu = \partial_\mu + i e A_\mu$, $F_{\mu \nu} = \partial_\mu A_\nu - \partial_\nu A_\mu$.
However, we will use a sextic potential
\begin{equation}
\label{eq:potential}
U(|\phi|^2) = |\phi|^2 (|\phi|^2 - 1)^2
\end{equation}
instead of a commonly used quartic potential. The vacua in this model form a union of a circle $\mathcal{V}_2$ and a point $\mathcal{V}_1$. More precisely, $\mathcal{V}_1 = \{ (\phi, A_\mu) : \phi = A_\mu = 0 \}$ and $\mathcal{V}_2 = \{ (\phi, A_\mu) : | \phi | = 1, A_\mu = 0 \}$. Due to this structure, topological defects (solitons) are present: our focus will be on domain walls, interpolating between $\mathcal{V}_1$ and $\mathcal{V}_2$. It is, however, worth mentioning, that the theory has also vortices $\mathcal{V}_2 \to \mathcal{V}_2$ with a winding number as a topological charge. The well-known Higgs mechanism is present in this model: at the vacuum $\mathcal{V}_2$, after fixing the phase of the scalar field, the vector field admits mass $m = \sqrt{2}e$.

We are interested in a special case of simple solutions, which can be derived from a restricted (1+1) dimensional model, obtained from the (3+1) dimensional theory with the following assumptions. First, let the gauge field have nonvanishing component only in the direction perpendicular to $x$, and denote it $A_\perp (x, t)$. Second, fix the phase of the scalar field, making it real, and assume that the scalar fields depends only on the $x$ coordinate and time, i.e. $\phi = \phi(x, t)$. Then, the full theory (\ref{eq:lagrangian}) reduces to a (1+1) dimensional restricted model
\begin{equation}
\label{eq:lagrangian_simplified}
\mathcal{L} = \frac12 \left( \partial_t A_\perp \right)^2 - \frac12 \left( \partial_x A_\perp \right)^2 + \left( \partial_t \phi \right)^2 - \left( \partial_x \phi \right)^2 - e^2 A_\perp^2 \phi^2 - U(\phi^2).
\end{equation}
The simplified equations of motion are
\begin{equation}
\begin{split}
\label{eq:EOM_simplified}
A_{\perp, tt} - A_{\perp, xx} + 2 e^2 A_\perp \phi^2 &= 0, \\
\phi_{tt} - \phi_{xx} + e^2 A_\perp^2 \phi + U'(\phi^2) \phi &= 0,
\end{split}
\end{equation}
where
\begin{equation}
\label{eq:potential_derivative_simplified}
U'(\phi^2) = 1 - 4 \phi^2 + 3 \phi^4.
\end{equation}
Looking for static solutions with $A_\perp = 0$, we obtain the second order ordinary differential equation (ODE)
\begin{equation}
    - \phi_{xx} + U'(\phi^2) \phi = 0
\end{equation}
with the known domain wall (kink) solution
\begin{equation}
\label{eq:kink}
\phi_D(x) = \sqrt{\frac{1 + \tanh x}{2}}.
\end{equation}
It is worth noting that the domain wall also obeys the first order ODE
\begin{equation}
\label{eq:BPS}
   \partial_x \phi_D = \sqrt{U \left(\phi_D^2 \right)}.
\end{equation}
Interactions of the static solutions with both scalar and vector fields can be studied. The interactions of the domain walls with a radiation in the scalar field without the gauge potential were studied in \citet{ROMANCZUKIEWICZ2017295}. In this paper, we focus on small waves in the gauge field instead.

Finally, we can derive the conservation laws in the simplified model. Using the Noether theorem, the energy-momentum tensor can be obtained:
\begin{equation}
\label{eq:energy_momentum_tensor}
\begin{split}
\tensor{T}{^0_0} &= \frac12 \left( \partial_t A_\perp \right)^2 + \frac12 \left( \partial_x A_\perp \right)^2
+ \left( \partial_t \phi \right)^2 + \left( \partial_x \phi \right)^2
+ e^2 A_\perp^2 \phi^2 + U\left(\phi^2\right)\,, \\
\tensor{T}{^1_1} &= -\frac12 \left( \partial_t A_\perp \right)^2 - \frac12 \left( \partial_x A_\perp \right)^2
- \left( \partial_t \phi \right)^2 - \left( \partial_x \phi \right)^2
+ e^2 A_\perp^2 \phi^2 + U\left(\phi^2\right)\,, \\
\tensor{T}{^0_1} &= -\tensor{T}{^1_0} = \partial_t A_\perp \, \partial_x A_\perp
+ 2 \, \partial_t \phi \, \partial_x \phi
\end{split}
\end{equation}
and the total energy and momentum are
\begin{equation}
\label{eq:total_energy_momentum}
E = \int_{-\infty}^{\infty}\tensor{T}{^0_0} {\rm d}x, \quad
P = -\int_{-\infty}^{\infty}\tensor{T}{^0_1} {\rm d}x.
\end{equation}
Then, from the conservation law
\begin{equation}
\label{eq:energy_momentum_conservation}
\partial_\mu \tensor{T}{^\mu_\nu} = 0
\end{equation}
we conclude that the change of the total momentum in time depends only on the asymptotic form of $\tensor{T}{^1_1}$, namely
\begin{equation}
\label{eq:energy_momentum_conservation_P_total}
\partial_t P = \left. \tensor{T}{^1_1} \right|_{-\infty}^\infty.
\end{equation}

\section{Linearized equations}
\label{sec:linearized}
In order to study the interactions of solitons with a radiation, we consider small perturbations around the domain wall:
\begin{equation}
\begin{split}
\phi(x, t) &= \phi_D(x) + b \, \delta \phi(x, t) \,, \\
A_\perp (x, t) &= a \, \delta A_\perp (x, t),
\end{split}
\end{equation}
where $a$ and $b$ are small real parameters. After substitution to (\ref{eq:EOM_simplified}), the equations for $\delta \phi(x, t)$ and $\delta A_\perp (x, t)$ in the leading order separate, and since we are interested only in interactions with gauge field waves in the linear regime, we put $b = 0$. Then, we can separate the variables:
\begin{equation}
\delta A_\perp (x, t) = \eta(x) e^{i \omega t} \,,
\end{equation}
obtaining a linear Schr\"odinger-like equation
\begin{equation}
\label{eq:linearized_equation_schroedinger}
-\eta''(x) + V(x) \eta(x) = \omega^2 \eta(x)
\end{equation}
with an effective potential
\begin{equation}
\label{eq:linearized_potential}
V(x) = e^2 \left( 1 + \tanh(x) \right).
\end{equation}

We are interested in the scattering modes, which asymptotically are plane waves propagating from the left to the right or vice versa. Such solutions have a wavenumber $\omega$ at $x \to -\infty$ and
\begin{equation}
\label{eq:wavenumber}
k = \sqrt{\omega^2 - 2e^2}
\end{equation}
at $x \to +\infty$. Therefore, we are looking for solutions $\eta_\rightarrow$ and $\eta_\leftarrow$, such that
\begin{equation}
\label{eq:S_matrix_definition}
\begin{split}
\eta_\rightarrow(x \to -\infty) &= e^{i x \omega} + S_{11} e^{-i x \omega}\,, \\
\eta_\rightarrow(x \to +\infty) &= S_{21} e^{i k x}\,, \\
\eta_\leftarrow(x \to -\infty) &= S_{12} e^{-i x \omega}\,, \\
\eta_\leftarrow(x \to +\infty) &= e^{-i k x} + S_{22} e^{i k x}\,,
\end{split}
\end{equation}
where  $e^{i \omega t} \eta_\rightarrow(x)$ describes a wave moving from $-\infty$ to $\infty$,  $e^{i \omega t} \eta_\leftarrow(x)$ is a wave moving in the opposite direction, and $S_{ij}$ are elements of the so-called $S$-matrix. Importantly, the waves at $+\infty$ are not propagating if $\omega^2 \leq 2e^2$. The wavenumber (\ref{eq:wavenumber}) can be understood as a gauge field acquiring mass $m = \sqrt{2} e$ at $|\phi| = 1$ vacuum due to the Higgs mechanism.

The solutions with the asymptotics discussed above can be found exactly:
\begin{equation}
\label{eq:linearized_solutions}
\begin{split}
\eta_\rightarrow(x) &= e^{i x \omega} \, _2F_1\left(-\frac{1}{2} i(k - \omega ), \frac{1}{2} i(k + \omega ); i\omega + 1; -e^{2x}\right) \\
&- \alpha  e^{-i x \omega} \, _2F_1\left(\frac{1}{2} i (k-\omega), -\frac{1}{2} i (k+\omega ); 1 - i\omega; -e^{2x}\right) \,, \\
\eta_\leftarrow(x) &= \beta  e^{-i x \omega} \, _2F_1\left(\frac{1}{2} i (k - \omega ), -\frac{1}{2} i (k + \omega ); 1 - i\omega; -e^{2x}\right) \,,
\end{split}
\end{equation}
where
\begin{equation}
\begin{split}
\alpha &= \frac{\Gamma (1+ i \omega) \Gamma \left(- \frac{i\omega}{2} - \frac{i k}{2} \right) \Gamma \left(- \frac{i \omega }{2} - \frac{ik}{2} + 1 \right)}
{\Gamma (1-i \omega ) \Gamma \left(\frac{i \omega }{2}-\frac{i k}{2}\right) \Gamma \left(\frac{i \omega}{2} - \frac{ik}{2} + 1 \right)} \,, \\
\beta &= 
\frac{\Gamma \left(-\frac{i k}{2}-\frac{i \omega }{2}\right) \Gamma \left(-\frac{i k}{2}-\frac{i \omega }{2}+1\right)}{\Gamma (-i k-2) \Gamma (1-i \omega ) \left(2e^2 + 3ik - \omega^2 + 2\right)} \,.
\end{split}
\end{equation}
This allows us to find the values of the $S$-matrix elements, and especially their moduli squared:
\begin{equation}
\label{eq:S_matrix_values}
\begin{split}
|S_{11}|^2 = |S_{22}|^2 &= \frac{\sinh^2 \left(\frac{1}{2} \pi \left(\omega -k\right)\right)}{\sinh^2 \left(\frac{1}{2} \pi \left(\omega + k\right)\right)} \, , \\
|S_{21}|^2 &= \frac{\omega}{k} (1 - |S_{11}|^2) \, , \\
|S_{12}|^2 &= \frac{k}{\omega} (1 - |S_{11}|^2),
\end{split}
\end{equation}
which will be used in the computation of the effective force exerted on the domain wall.

\section{Effective dynamics}
\label{sec:dynamics}

To study the domain wall dynamics for low velocities, the wall can be approximated as a Newtonian particle. More precisely, we assume that its shape remains constant and treat its center position $x_0$ as a function of time, obeying
\begin{equation}
M \ddot{x}_0(t) = F.
\end{equation}
Let $P_v$ be the total momentum of the boosted domain wall with a velocity $v$. Using equation (\ref{eq:total_energy_momentum}), we obtain the effective mass:
\begin{equation}
\label{eq:soliton_mass}
M = \left. \frac{\partial P_v}{\partial v} \right|_{v=0} = \frac12.
\end{equation}
The effective force is
\begin{equation}
\label{eq:effective_force}
F = \langle \partial_t P \rangle_T\,,
\end{equation}
where $P$ is the total momentum of the domain wall and the scattered wave, and $\langle \cdot\rangle_T$ means the average over the period of the oscillations.

Now we can present the specific results describing the interaction between the domain wall and the radiation. We assume that a sinusoidal plane wave of the form
\begin{equation}
\label{eq:wave_sinusoidal}
\delta A_{\perp}(x, t) = \frac{a}{2i} \left( e^{- i \omega t} \eta(x) - c.c. \right)
\end{equation}
is scattered on a soliton, when $\eta$ is the asymptotic form of $\eta_\rightarrow$ or $\eta_\leftarrow$ from the equation (\ref{eq:S_matrix_definition}) and the amplitude $a$ is sufficiently small, in order to use the linear regime. Then, applying equation for the time derivative of the total momentum (\ref{eq:energy_momentum_conservation_P_total}) and the formula for the effective force (\ref{eq:effective_force}) to the setup consisting of a domain wall and the sinusoidal wave (\ref{eq:wave_sinusoidal}), the respective effective forces are obtained:
\begin{equation}
\label{eq:force}
\begin{split}
F_\rightarrow &= \frac{1}{2} a^2 \left[\omega^2 \left( 1 + \left| S_{11} \right|^2 \right) - k^2 \left| S_{21} \right|^2 \right]\,, \\
F_\leftarrow &= \frac{1}{2} a^2 \left[ -k^2 \left( 1 + \left| S_{22} \right|^2 \right) + \omega^2 \left| S_{12} \right|^2 \right]\,.
\end{split}
\end{equation}
Substituting the explicit formulas (\ref{eq:S_matrix_values}), we get
\begin{equation}
\label{eq:force_explicit}
\begin{split}
F_\rightarrow &= a^2 \omega \frac{-k \sinh (\pi \omega) \sinh \left(\pi  k \right)+\omega  \cosh (\pi \omega ) \cosh \left(\pi  k \right) - \omega}{\cosh \left(\pi \left(k + \omega \right)\right) - 1}\,, \\
F_\leftarrow &= a^2 k \frac{\omega \sinh (\pi \omega) \sinh \left(\pi k \right) - k \cosh (\pi \omega) \cosh \left(\pi k \right) + k}{\cosh \left(\pi \left(k + \omega \right)\right) - 1}\,.
\end{split}
\end{equation}
Both forces are always positive (or zero), regardless of the direction of the incoming wave. This means that if the wave is coming from the left, we observe the positive radiation pressure (PRP) and when it is coming from the right, we see the negative radiation pressure (NRP). For $\omega \to \infty$:
\begin{equation}
F_\rightarrow = F_\leftarrow = \frac{a^2 e^2}{2},
\end{equation}
which is not surprising, since then also $k \to \omega$ and the waves on both sides become more similar to each other. For $\omega \to \sqrt{2} e$ waves with different directions behave differently:
\begin{equation}
\begin{split}
F_\rightarrow &= 2 a^2 e^2, \\
F_\leftarrow &= 0. \\
\end{split}
\end{equation}

\section{Numerical results}
\label{sec:results}
To check the validity of the approximate effective forces (\ref{eq:force_explicit}), the equations (\ref{eq:EOM_simplified}) were solved numerically, and the results were compared. The initial condition was the domain wall $\phi = \phi_D$ and the `wave train'
\begin{equation}
    A_\perp(x, t) = a \sin(\omega x - \omega t) \frac{\tanh(x - 20 + L - v_1 t) - \tanh(x + 20 - v_1 t)}{2}
\end{equation}
for the wave going from left to right or
\begin{equation}
    A_\perp(x, t) = a \sin(-k x - \omega t) \frac{\tanh(x - 20 + v_2 t) - \tanh(x + 20 - L + v_2 t)}{2}
\end{equation}
for the wave propagating from right to left, where $v_1 = 1$, $v_2 = k/\omega$ are group velocities and $x \in [-L, L]$ with $L = 250$. The RK4 method was used, with space and time steps $\Delta x = 0.01$ and $\Delta t = 0.005$ respectively. Accelerations were measured by fitting the quadratic function to the position of the soliton in the time range $20$ to $200$. 

\begin{figure}
	\includegraphics[width=\columnwidth]{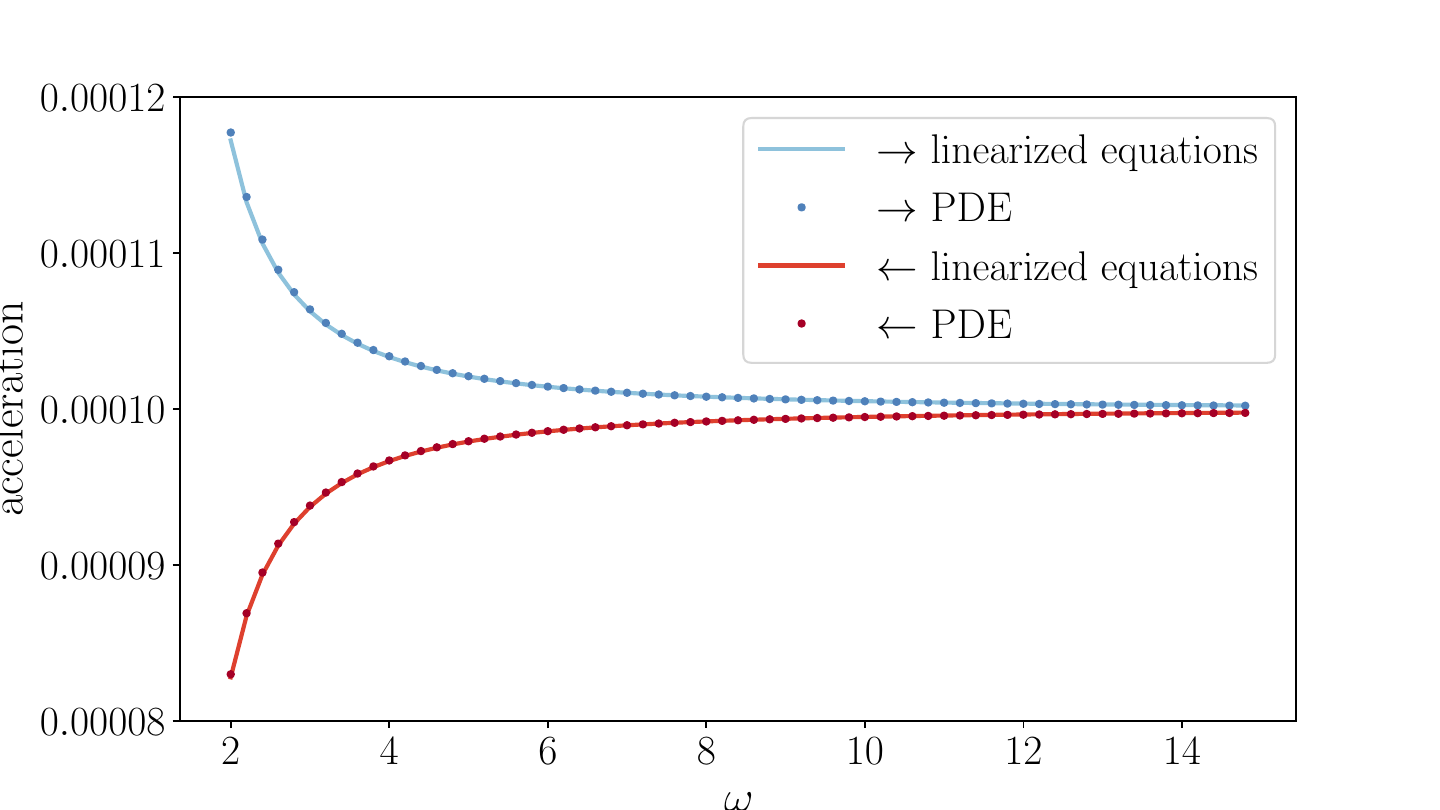}
	\caption{Acceleration of the domain wall interacting with a wave coming from the left (upper, blue) or right (lower, red) with an amplitude $a = 0.01$, a charge $e = 1$ and different frequencies.}
	\label{fig:frequency}
\end{figure}

\begin{figure}
	\includegraphics[width=\columnwidth]{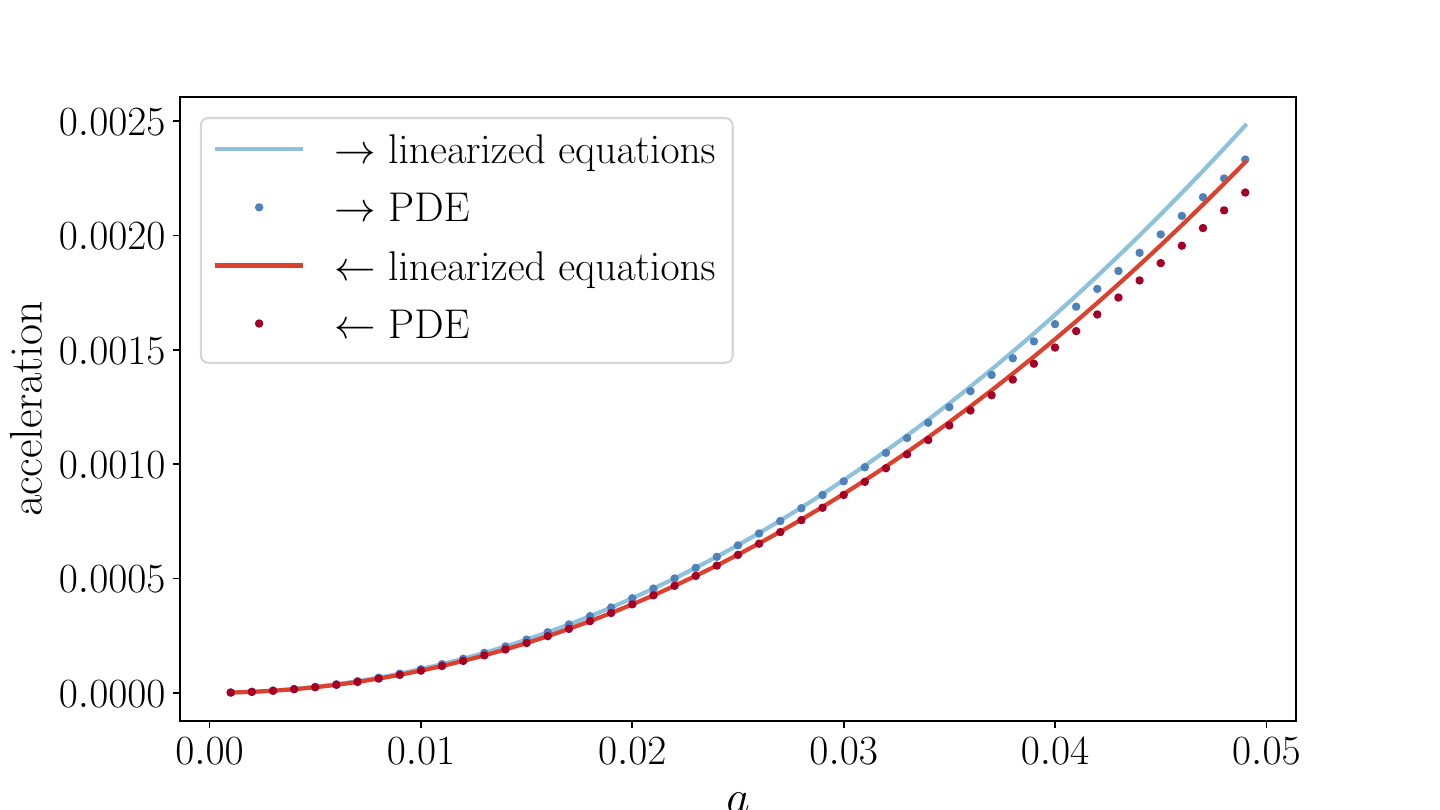}
	\caption{Acceleration of the domain wall interacting with a wave coming from the left (upper, blue) or right (lower, red) with a frequency $\omega = 4$, a charge $e = 1$ and different amplitudes.}
	\label{fig:amplitude}
\end{figure}

\begin{figure}
	\includegraphics[width=\columnwidth]{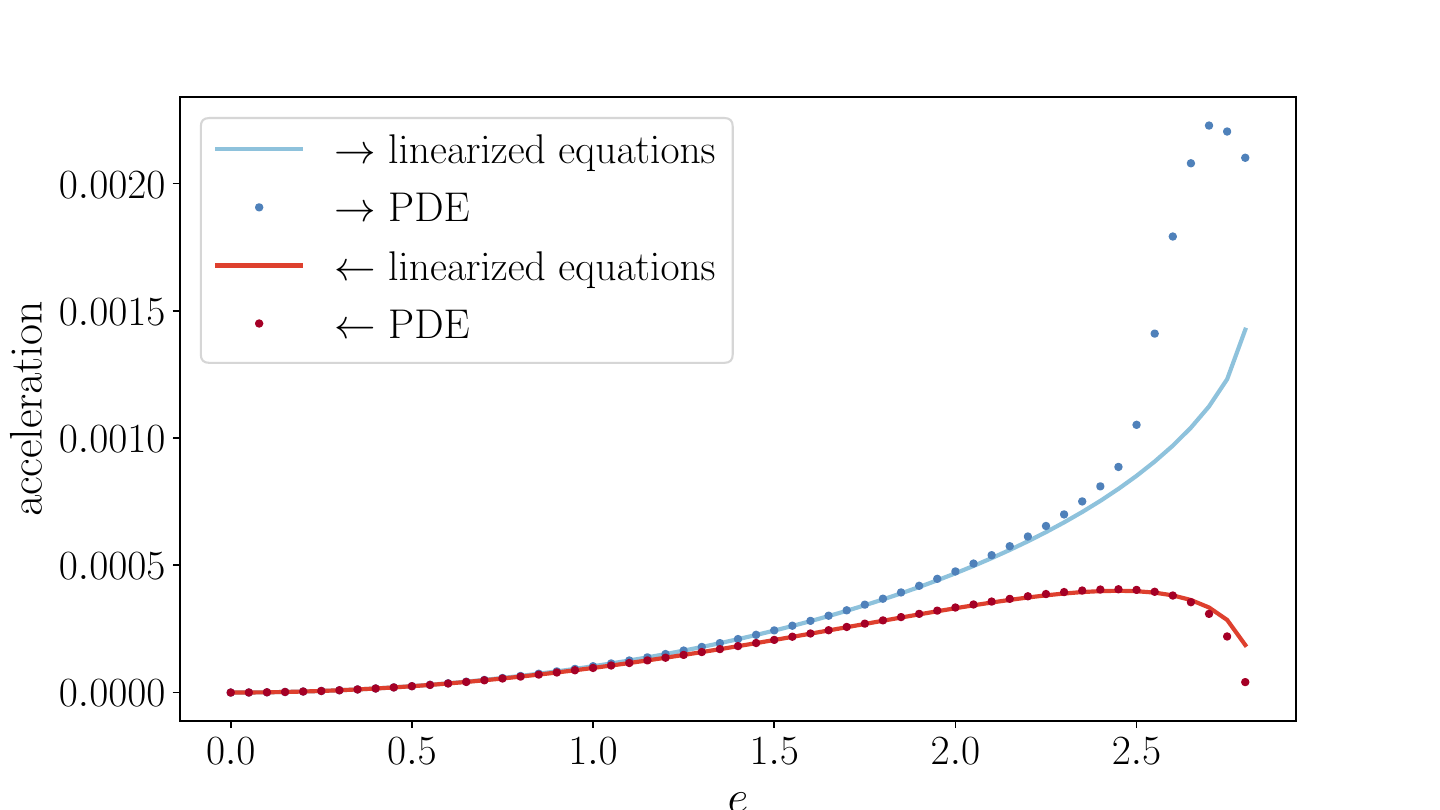}
	\caption{Acceleration of the domain wall interacting with a wave coming from the left (upper, blue) or right (lower, red) with an amplitude $a = 0.01$, a frequency $\omega = 4$ and different charges $e$.}
	\label{fig:charge}
\end{figure}

\begin{figure}
	\includegraphics[width=\columnwidth]{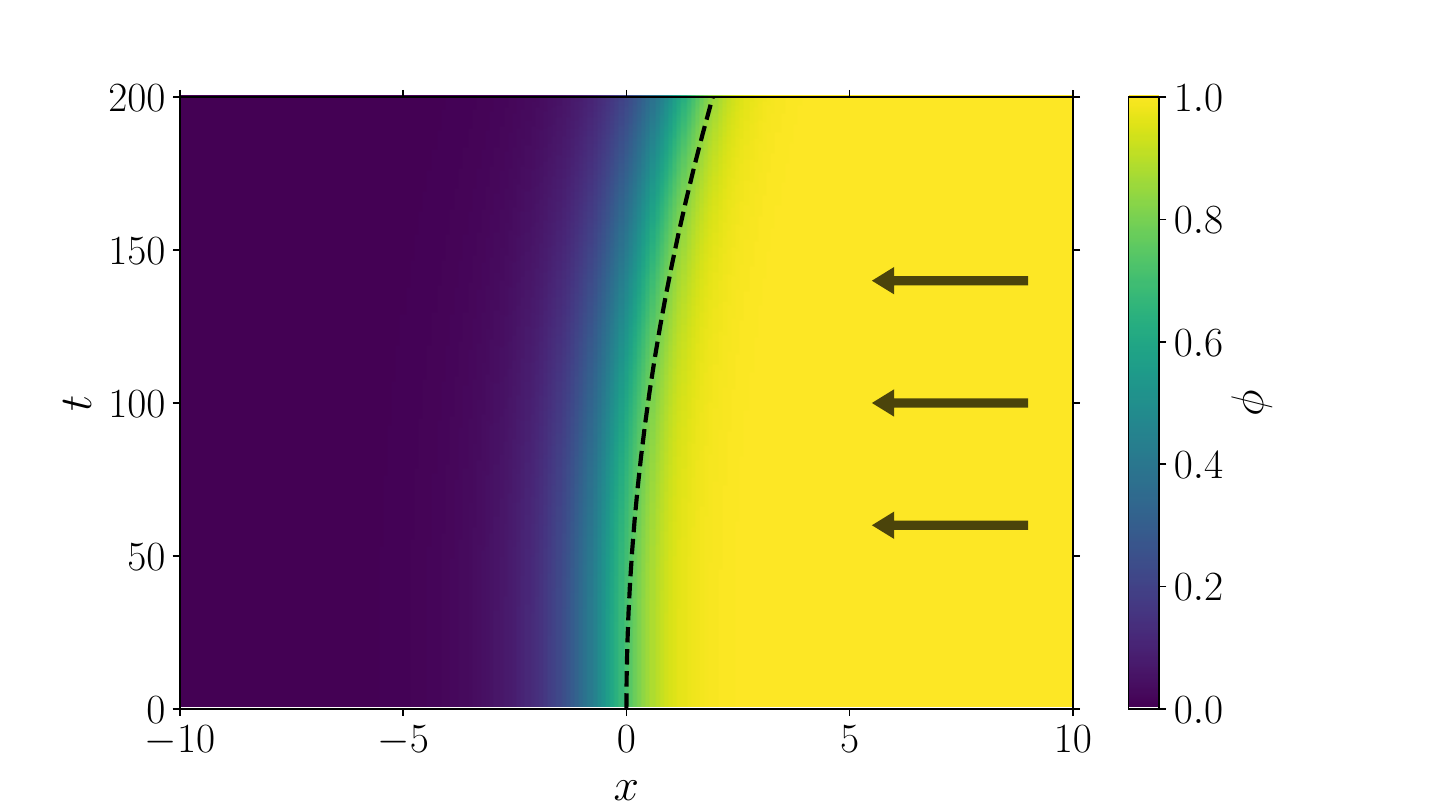}
        \includegraphics[width=\columnwidth]{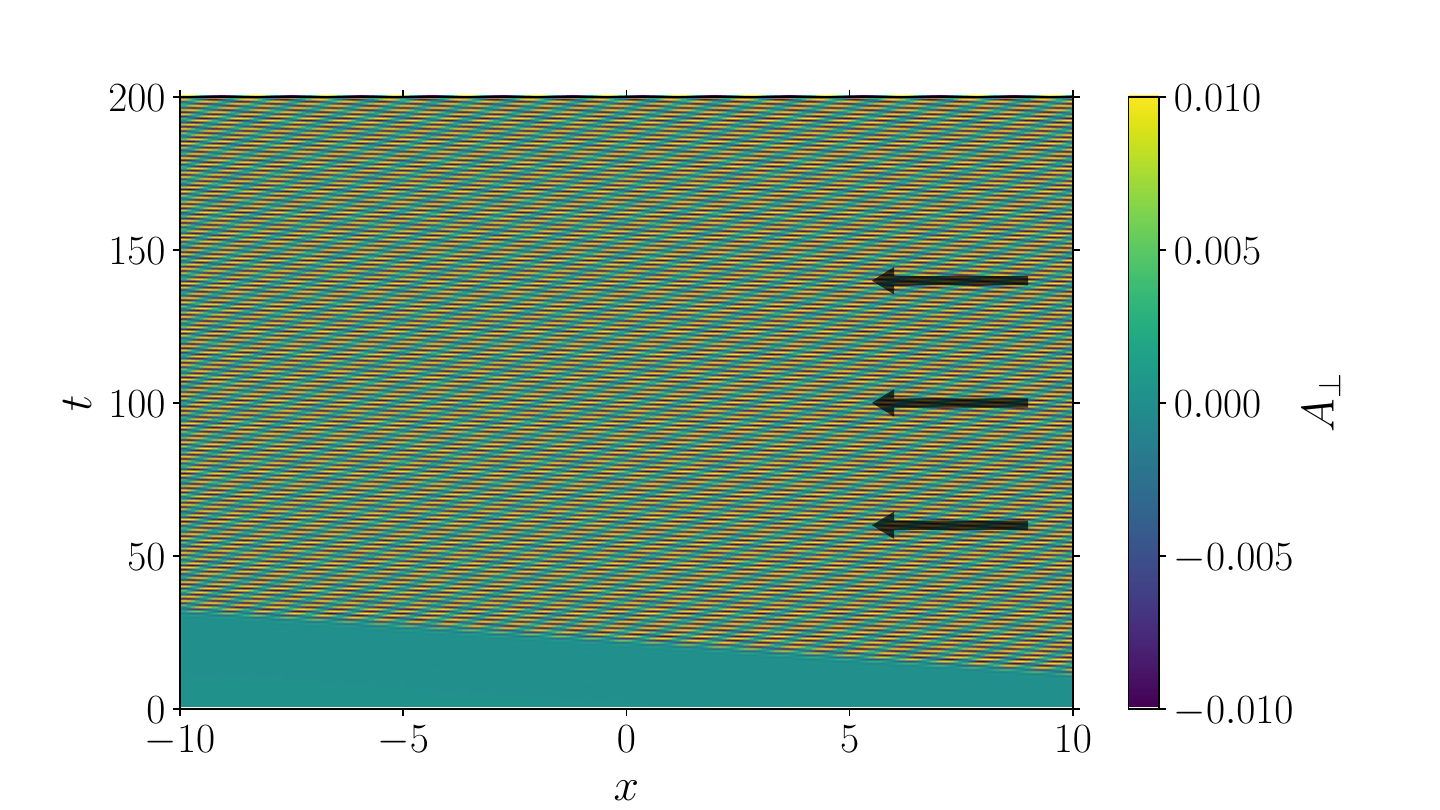}
	\caption{Example of an evolution of the domain wall interacting with a gauge field wave coming from the right with an amplitude $a = 0.01$, a frequency $\omega = 4$ and a charge $e = 1$. Both scalar field (upper) and gauge field (lower) are presented. The dashed line represents the soliton trajectory with acceleration obtained from the linearized approximation. Arrows mark the direction of the incoming wave.}
	\label{fig:example}
\end{figure}

The results, presented in figures \ref{fig:frequency}, \ref{fig:amplitude} and \ref{fig:charge}, agree well with the approximate formulas derived from the linearized equations. An example of the soliton evolution with NRP can be seen in the figure \ref{fig:example}. As expected, the accuracy of the approximation drops for larger amplitudes (Fig. \ref{fig:amplitude}), and charges (Fig. \ref{fig:charge}), due to the nonlinear effects.

\section{Conclusions}
Using the linearized Newtonian approximation and numerical simulations, we have shown that the domain walls always accelerate towards $\phi = 0$ vacuum. Therefore, we observe both PRP and NRP, depending on the direction of the incoming wave. This behavior is very similar to the one observed in the $\phi^6$ model without a gauge field \citep{ROMANCZUKIEWICZ2017295}, and we expect that it would play a similar role in multikink collisions, although it requires further study.

Another possibility for further research would be to study oscillons and their interactions with kinks. Oscillons can play an important role in many physical processes and have interesting, nontrivial dynamics, and moreover, they can store enough energy to bounce back kinks \citep{Dorey2023}. Furthermore, radiation pressure exerted on more complicated topological solitons, i.e. vortices, can be studied in this model.

Observation of negative radiation pressure in (1+1) abelian Higgs model with a sextic potential is yet another indication that this phenomenon is widespread in theoretical physics. Although the results of this paper are not applicable to experiments directly, the studied system can be understood as an important simple toy model, expanding the mathematical tools needed to find NRP in nature.

\section*{Acknowledgements}
This work has been supported by the Priority Research Area under the program Excellence
Initiative – Research University at the Jagiellonian University in Krak\'ow. The author would also like to express the gratitude to Tomasz Romańczukiewicz for useful discussions.

%-------------------------------------------------------------------
% bibliography
%-------------------------------------------------------------------
%\bibliographystyle{elsarticle-harv}
\bibliographystyle{unsrt}
\bibliography{phi6_gauge_bibliography}

\end{document}